# Optical read and write of spin states in organic diradicals

Rituparno Chowdhury[1], Petri Murto[1,2], Naitik A. Panjwani[3], Yan Sun[4], Pratyush Ghosh[1], Yorrick Boeije[1], Vadim Derkach[4,5], Seung-Je Woo[1], Oliver Millington[2], Daniel G. Congrave[2], Yao Fu[1,2], Tarig B. E. Mustafa[1,2], Miguel Monteverde[4], Jesús Cerdá [6], Jan Behrends[3], Akshay Rao[1], David Beljonne[6], Alexei Chepelianskii[4]*, Hugo Bronstein[2]*, Richard H. Friend[1]*

[1]The Cavendish Laboratory, University of Cambridge, J. J. Thomson Avenue, Cambridge CB3 0HE, United Kingdom.

[2]Yusuf Hamied Department of Chemistry, University of Cambridge, Lensfield Road, Cambridge CB2 1EW, United Kingdom.

[3]Berlin Joint EPR Lab, Fachbereich Physik, Freie Universität Berlin, 14195 Berlin, Germany.

[4]LPS, Université Paris-Saclay, CNRS, UMR 8502, Orsay F-91405, France.

[5]O. Ya. Usikov Institute for Radiophysics and Electronics of NAS of Ukraine 12, Acad. Proskury st., Kharkiv, 61085, Ukraine.

[6]Laboratory for Chemistry of Novel Materials, University of Mons, 7000, Mons, Belgium.

*email: rhf10@cam.ac.uk, hab60@cam.ac.uk, alexei-chepelianskii@universite-paris-saclay.fr

## Abstract

Optical control and read-out of the ground state spin structure has been demonstrated for defect states in crystalline semiconductors, including the diamond $NV^-$ center, and these are promising systems for quantum technologies. Molecular organic semiconductors offer synthetic control of spin placement, in contrast to current limitations in these crystalline systems. Here we report the discovery of spin-optical addressability in a diradical molecule that comprises two trityl radical groups coupled via a fluorene bridge. We demonstrate the three important properties that enable operation as a spin-photon interface: (i) triplet and singlet spin states show photoluminescence peaked at 640 and 700 nm respectively; this allows easy optical measurement of ground state spin. (ii) the ground state spin exchange is small (~60 µeV) that allows preparation of ground state spin population. This can be achieved by spin-selective excited state intersystem crossing, and we report up to 8% microwave-driven contrast in photoluminescence. (iii) both singlet and triplet manifolds have near-unity photoluminescence quantum yield, which is in contrast to the near-zero quantum yields in prior reports of molecular diradicals. Our results establish these tuneable open-shell organic molecules as a platform to engineer tailor-made spin-optical interfaces.

**Main**

Optical initialisation and read-out of spin states has been of growing interest to the emerging field of quantum technologies with implications in quantum computing(*1, 2*), communication(*3*), teleportation(*4*) and sensing(*5*). This spin-optical interface and its applications have been demonstrated through solid-state defects(*6–8*). However synthetic control over the optical-spin and spin-spin interaction as well as controlled fabrication of multi-spin architectures remain challenging. Organic synthesis provides an attractive route to achieve this, through its capability to provide atomistic control over molecules, which could provide control over the placement of quantum spin-optical units precisely at well-defined distances and in large assemblies. Since these properties are present at the molecular level, there would be relative host independence further simplifying implementation in devices.

In molecular organo-metallic systems it is still possible to create an open-shell, non-zero net spin, ground state and this has been explored to successfully create spin-optical interfaces(*9*). Fully organic systems offer geometric and spatial control through readily accessible synthetic chemistry methods. There are reports of coherent control over photoexcited high spin states(*10–12*), spin-singlet based quantum optics(*14*), and long coherence times(*15*), but these are from non-luminescent systems. Optical read-out following excited state spin manipulation has recently been reported(*13*). Recent theoretical studies identify alternant diradicals as candidates to construct organic spin-optical qubits(*17, 18*). In alternant diradicals the atoms contributing to the π-framework can be divided into two classes such that atoms within each class are never adjacent to each other. However, alternant diradical systems reported till date, much like mono-radical systems, suffer from a symmetry forbidden dark transition(*16*). This has prevented diradical ground states from being optically prepared or manipulated, though it has been shown under dark conditions that the very long lived coherent superpositions of the ground state can be achieved using microwave pulses(*15*). Discovering an open-shell, highly luminescent, magnetic molecule capable of supporting a spin-optical interface to control the ground-state spin populations would provide transformative new opportunities towards the design of on-demand quantum information units with precise spin-spin and spin-optical control.

Recently, we and others have shown that carbon-centred chlorinated trityl mono-radicals, TTMs, coupled to electron donors such as carbazole provide a family of efficient luminescent radicals (*19–23*) that have been of growing interest and show promise as light emitters in LEDs. For these it is possible to engineer an emissive paramagnetic spin-doublet to be the lowest excited state (*22, 23*), obviating the challenge of dark triplet states found in closed shell systems currently used in OLEDs. The stability of the TTM radical was increased through mesitylation(*24*) which breaks excited state electronic symmetry and thus brightens the first $D_0 \rightarrow D_1$ transition(*25*), this led to the discovery of $M_2TTM$ based doublet emitters which can be easily incorporated into a range of synthetically accessible molecular scaffolds.

Here, we report the discovery of an organic diradical spin-optical qubit where two $M_2TTM$ radicals are bridged with a molecular unit that is selected to act as bridge to the trityl radicals that maintains alternant symmetry. As we report below, this system, surprisingly, shows near-unity Photoluminescence Quantum Yield, PLQY, while allowing coherent manipulation of the ground-state spin-population. The discovery of such an open-shell, luminescent, diradical molecule that is capable of supporting a spin-optical interface to control the ground-state spin populations provides transformative new opportunities towards the design of on-demand quantum information units with precise spin-spin and spin-optical control.

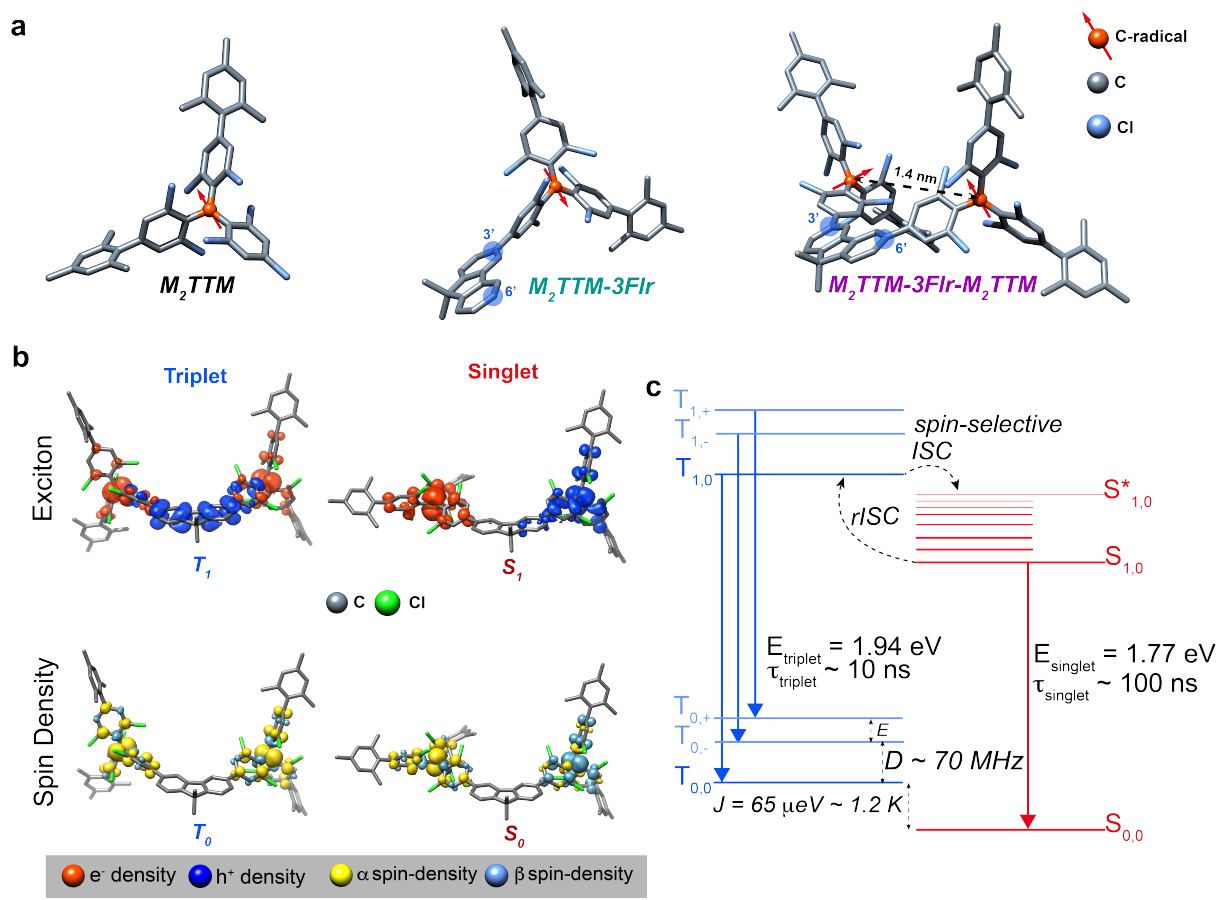

*Figure 1: Molecular design of luminescent diradical spin-optical units. (a) Molecular structure obtained from single-crystal X-ray crystallography of the M$_2$TTM radical and M$_2$TTM-3Flr radical and the geometry optimized structure for M$_2$TTM-3Flr-M$_2$TTM diradical. The substitution positions in the fluorene linker are highlighted alongside the spin-spin distance in the diradical. (b) Illustrations of the exciton wavefunction and spin density in the excited and ground states of the fluorene substituted diradical M$_2$TTM-3Flr-M$_2$TTM. A triplet exciton is shown for a diradical geometry for which the exciton extends across the whole molecule, illustrating the HOMO to SOMO excitation; note that for this alternant hydrocarbon there will be a corresponding, degenerate, SOMO to LUMO component to the excitation. The relaxed singlet exciton is shown for a twisted geometry, showing charge transfer between the SOMO states to form a zwitterionic excitation. (see SI Section X for computational methods). (c) Energy level diagram for ground and excited states of the singlet and triplets, shown at close to zero magnetic field. Energy values as determined from experiment are indicated.*

Figure 1(a) shows the structure of the molecule of interest. We have coupled two radical units, M$_2$TTM, via 3,6-substitution on a dioctylfluorene, Flr, linker. Note that this 3,6 coupling maintains a diradical ground state, and does not allow full π-conjugation between the radical groups. The monoradical is labelled M$_2$TTM-3Flr and the diradical is labelled M$_2$TTM-3Flr-M$_2$TTM. This diradical shows close to unity photoluminescence yield, Photoluminescence Quantum Yield, PLQY, from a triplet-to-triplet band near 640 nm and from a singlet-to-singlet band near 700nm. (Structures and photophysical properties of similar diradical systems are shown in Supplementary section S1).

Measurements presented here were carried out in dilute solutions (100 µM solutions of toluene) or in spin-coated thin films of polystyrene where the diradical concentration was ≤0.1wt%. Polystyrene serves as a non-polar host in which the diradical can be dispersed uniformly. We find it necessary to keep these diradical molecules well separated through dilution to prevent intermolecular energy and spin transfers. At these concentrations, aggregation effects are unimportant, see SI-Section.XII. Samples for PLDMR were prepared by doping diradicals at 1 nM concentration into 1,3,5-trichlorobenzene crystals.

| Molecule | $\lambda_{abs,onset}$ (nm) | $\lambda_{PL}$ (nm) | $\Phi_{PL}$ (%) | $\tau_{PL}$ (ns) |
|---|---|---|---|---|
| **M$_2$TTM** | 580 | 595 | 12 | 21 |
| **M$_2$TTM-3Flr** | 620 | 605 | 9 | 9 |
| **M$_2$TTM-3Flr-M$_2$TTM** | 620 | 635 | 92 | 10(52%), 100(48%) |

*Table 1: Optical properties of the unsubstituted monoradical M$_2$TTM, the 3Flr substituted monoradical M$_2$TTM-3Flr and the 3Flr substituted diradical M$_2$TTM-3Flr-M$_2$TTM.*

Figure 2(a) shows the absorption end PL spectrum of 100 µM toluene solutions of M$_2$TTM, M$_2$TTM-3Flr, and M$_2$TTM-3Flr-M$_2$TTM, with properties summarised in Table 1. The optical absorption spectra of M$_2$TTM shows a first absorption peak at 590 nm, which redshifts for M$_2$TTM-3Flr to 620 nm, but does not shift further for the diradical. M$_2$TTM exhibits PL at 590 nm in solution with a PLQE of 12% (*26*). For the substituted monoradical M$_2$TTM-3Flr the PL undergoes a red-shift to 605 nm, and now has strong vibronic features at 640 nm and 705 nm but also has a lower PLQY of 9%. Radiative lifetimes, related to the measured lifetime, $\tau_{PL}$, through the equation $\tau_{PL}$/PLQY, are 175 ns for M$_2$TTM and 100 ns for the M$_2$TTM-3Flr monoradical. We note that M$_2$TTM-3Flr, like the M$_2$TTM radical, is an alternate molecule. This alternancy symmetry imposes particle-hole symmetry (*27, 28*) which causes the lowest absorption band to have equal and opposing contributions from HOMO$_{Flr}$→SOMO$_{M2TTM}$ and SOMO$_{M2TTM}$→LUMO$_{Flr}$ transitions, causing it to be substantially forbidden. Thus, the absorption strength and the radiative lifetime for the M$_2$TTM-3Flr monoradical are little changed from the M$_2$TTM radical (*25*). This in contrast to carbazole systems where alternancy symmetry is broken, with the carbazole acting as electron donor where the particle-hole symmetry is broken which brightens the D$_0$→D$_1$ transition(*29, 30*). The diradical shows significantly red-shifted PL with peak emission at 640 nm and enhanced PLQE to above 90%. We note that there is a second peak in the PL near 700 nm, which, as developed later, is not a vibronic overtone, but is due to a spin-singlet excited state, while the 640 nm feature is due to a spin triplet excited state.

Figures 2(b) shows time resolved PL, TRPL, measurements at 200 K on the 0.1wt% doped polystyrene films, see also Table 1. Though the M$_2$TTM, M$_2$TTM-3Flr show fast mono-exponential decay (see Figure S8), but the M$_2$TTM-3Flr-M$_2$TTM diradical shows greatly slowed kinetics; about half of the of the emission, peaked at 640 nm, has a fast lifetime of 10 ns while the remaining half of the emission, peaked near 700 nm, exhibits a significantly delayed lifetime of 100ns and beyond. We see that this slow emission is responsible for the red region of the steady state spectrum (Figure S9). PL measurements at other temperatures are reported in figures S10-12. Though the initial decay of the 640 nm feature is fast (10 ns) it shows a temperature dependent long-lived component, and this slow component has an activation energy around 10 meV (Figure S.12). We consider the slow component of the 640 nm emission to be due to the reverse-intersystem crossing (rISC).

Figure 2(c) shows the temperature dependent PL intensity in 0.1wt% spin-coated thin-films of the 640 nm and 700nm peaks in the 300 K – 0.6 K regime. The 700 nm PL increases while the 640 nm PL decreases at lower temperatures, though the integrated-PL intensity does not change (Figure S.14). The slope of the curves changes twice: First at ~120 K and then again at ~1.0 K, correlating from the TRPL kinetics we can assign the initial slope-change event to the thermal energy of the system falling below the rISC activation energy and second, we show later, is when it falls below the antiferromagnetic exchange energy of the diradical.

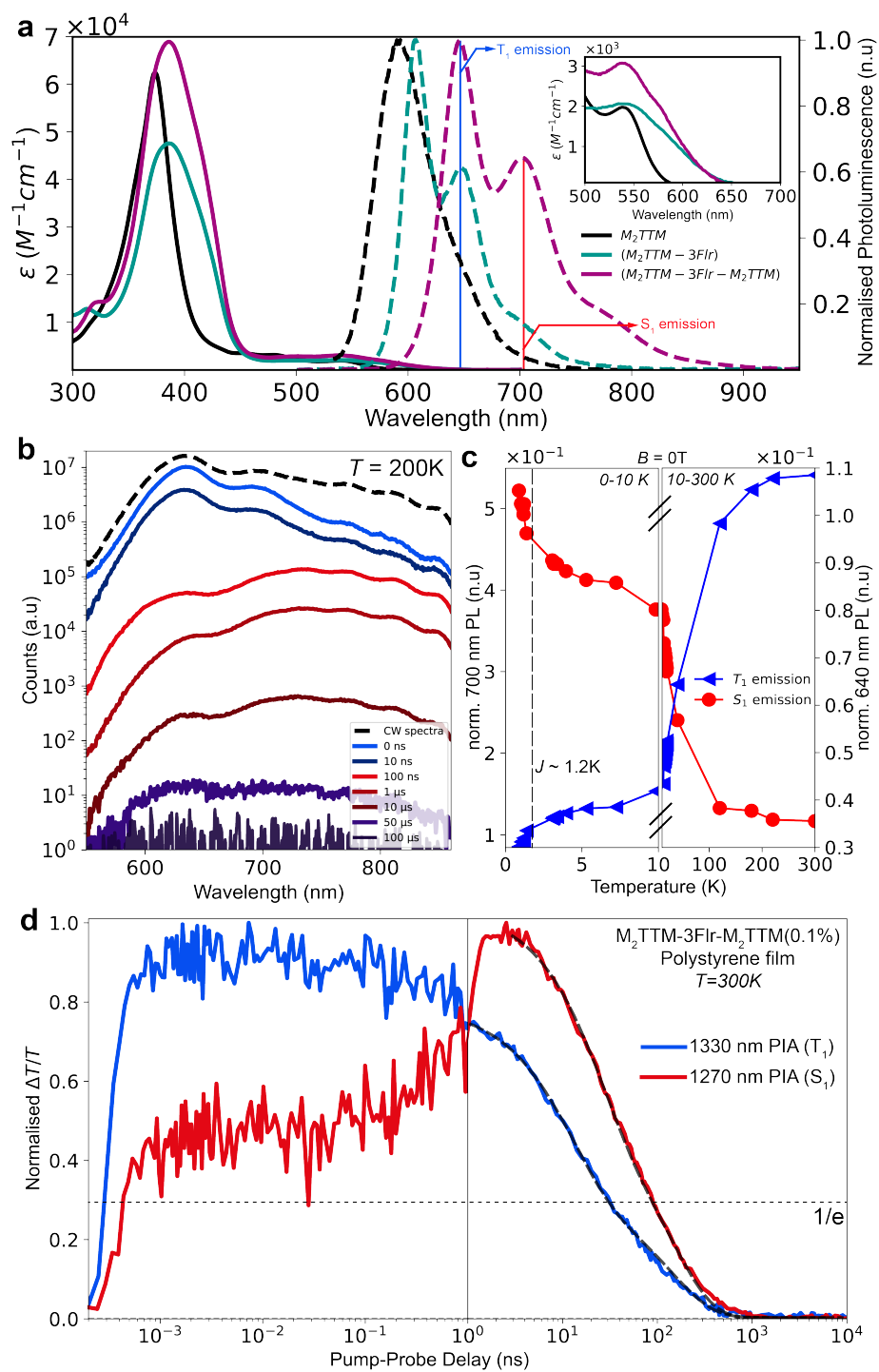

*Figure 2: **Photophysics of the luminescent diradical.** **(a)** The absorption and photoluminescence spectra are presented for the unsubstituted monoradical $M_2TTM$ (in black), the fluorene-substituted monoradical $M_2TTM$-3Flr (in green), and the fluorene-substituted diradical $M_2TTM$-3Flr-$M_2TTM$ (in purple). These spectra were acquired under ambient conditions following a 532 nm excitation in a 100 µM toluene solution. The **inset** shows absorption from 500 – 700 nm. **(b)** Time-resolved and CW PL for the $M_2TTM$-3Flr-$M_2TTM$ diradical at 200 K. **(c)** Zero-field Temperature dependent PL at 640 nm (blue triangles) and 700 nm (red circles) from 300 K to 0.6 K. The break point in the graph is 10 K. **(d)** Kinetic traces of the photoinduced absorption (PIA) features at 520 nm/1330 nm (blue) and at 610 nm/1270 nm (red) from the transient absorption spectra and the associated fits (dashed lines) for a multiexponential decay model. The exponential decay constants and weights are quoted in the figure. The data were obtained from a $M_2TTM$-3Flr-$M_2TTM$(0.1%):Polystyrene doped polystyrene film following a 532 nm excitation pulse at a fluence of 6 µJcm$^{-2}$.*

Figure 2(d) shows kinetic traces of the main peaks obtained from transient optical absorption spectroscopy. A detailed report and analysis of Transient optical absorption studies is provided in SI-Section.VI. In summary, induced absorption bands near 520 nm (2.05 eV) and 1330 nm (0.93 eV) are associated with initial charge transfer, to form the $M_2TTM$ anion(22), and bands at 610 nm (2.28 eV) and 1270 nm (0.98 eV) build up over the first 3 ns, as shown in Figure 2(c). We associate these two features with the kinetics of the 640 nm and 700 nm emissions respectively. There is evidence for an earlier time excited state evolution within the first 100 ps, which may be related to geometrical relaxation in the excited state, see SI-Section.VI.

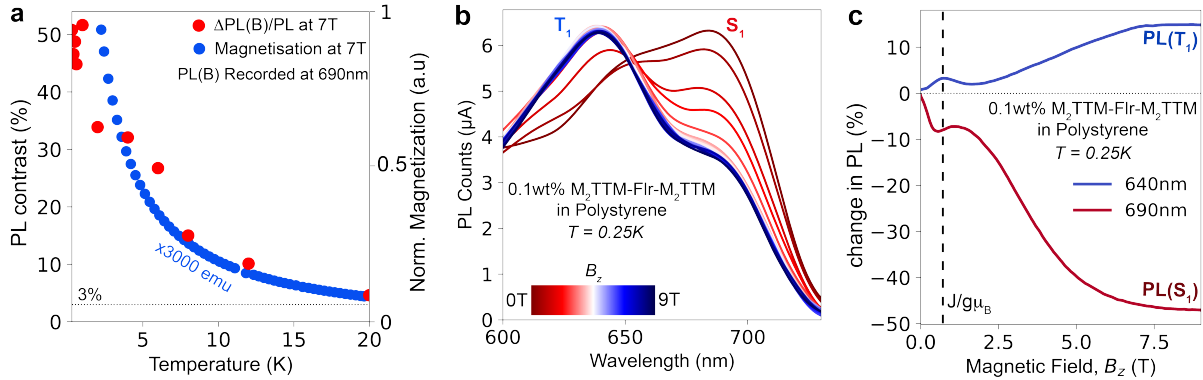

*Figure 3: Magneto-optic interface in luminescent diradicals. (a) Temperature dependence of the MPL contrast at 7T (red) for the 690nm emission compared to the temperature dependence of the magnetization at 7T (blue) measured using SQUID magnetometry. We investigate the temperature range from 0.3K to 12K and magnetization measured using SQUID magnetometry (b) PL spectrum recorded while increasing the perpendicular magnetic field applied to the sample plane. The spectra are obtained at 0.25K for a 0.1wt% doped polystyrene films. (c) PL intensity versus the magnetic field. One can smoothly vary this population using the applied B-field which can be read out using the relative intensity of the 640 nm and 700 nm PL. The data were obtained from a $M_2TTM$-3Flr-$M_2TTM$(0.1%):Polystyrene doped polystyrene film by using a 400 nm laser excitation at <1 $\mu Jcm^{-2}$ was used to pump the sample.*

Figure 3 shows the magneto-optical studies on spin-coated 0.1wt% thin-films of the diradical. We find that the 640 nm and 700 nm emission feature are strongly dependent on applied magnetic field. In the ground state the two spins, localised on each of the $M_2TTMs$, are weakly antiferromagnetically coupled, as shown in the temperature dependent magnetisation curves shown in Figure 3(a), with an antiferromagnetic exchange energy near 1K and this is also captured using PL through the rise in 700 nm and fall in 640 nm emission below 1.2 K as shown in Figure 2(a). Figure 3(b) shows the continuous wave PL spectra, recorded at 0.25 K at magnetic fields up to 9 T. In summary, the 700 nm emission dominates at zero-field, where the singlet state is occupied, and at higher fields, where the Zeeman energy exceeds the exchange energy, the 640 nm emission dominates. We associate this with a magnetic spin triplet ground state. Figure 3(c) shows evidence for a resonance where the Zeeman energy equals the ground state exchange energy, and is seen as expected near 0.65 T. In the region close to exchange, we obtain PL spectra that are linear combinations of the singlet and triplet PL which are proportional their relative populations (see Figure S.19).

In Figure 3(d) we show the model that can give rise to this field-dependent behaviour, more details are provided in the SI, see figures S21-23. We find that the field and temperature-dependent PL quantitatively matches the magnetisation measurements, as shown in Figure 3(a). This establishes that the two emissions are associated with the magnetization of the sample arising from the two possible overall spin states, singlet at 700 nm and triplet at 640 nm. Clearly, the red shift from triplet to singlet emission, of 170 meV, is a vastly higher energy than the ground state exchange and reveals a much increased 'effective' exchange energy in the excited state. We are also able to match results from the model for the magnetic field behaviour to the observed transient photophysical data (SI Section IV.A and IV.B).

The experimental observations set out in Figures 2 and 3 present a very clear picture: in the ground state the two spins are weakly coupled, but the energetics of the photoexcited states are strongly affected by the overall spin state. With the experimental observations in mind, we propose the following overall photophysical model, illustrated in Figure 1(b). We can assume that the initial photoexcitation occurs via fluorene HOMO/LUMO to SOMO transitions due to the similarity of the absorption spectrum of the diradical to the monoradical (See Figure 2(a)). These new states can be of either singlet or triplet multiplicity, denoted as $S_1$ and $T_1$ respectively. Extension of the exciton to both radical units will be dependent on molecular geometry and Figure 1(b) illustrates an extended triplet and a singlet with zwitterionic character. Following the initial photoexcitation the Frank-Condon states should undergo geometrical relaxation leading to excited state symmetry-breaking which enables the red-shifted and faster PL for triplet emission(*30*). In the case of the $S_1$ state this can be further energetically stabilised under a dielectric field with $\varepsilon$ > 2.2 (See SI Section X) by acquiring zwitterionic character through undergoing a second charge-transfer from the remaining $M_2TTM$ radical to the fluorene bridge. This scenario is active in our studies as the dielectric constant of polystyrene is 2.6 and that of toluene is 2.3. This decouples electron and hole, and leads to the slow radiative emission rate measured for the 700 nm emission from the singlet state. We note that diradicals with bright zwitterionic states have recently been reported(*31, 32*). The unique feature of our studied material is the dual fluorescence from both the triplet and singlet excited states with near unity quantum yield. We are unaware of any other organic material that shares these spin optical properties.

The origin of high PLQEs in some of the radical donor-acceptor systems has been explored by Ghosh *et al*(*33*). Generally, PLQEs are low for red and IR emitting molecular semiconductors due to the multiphonon decay assisted non-radiative loss especially when the electronic coupling to high frequency vibrational modes is strong. Ghosh *et al.* show two mechanisms can deactivate this vibrational coupling. The first is where the electronic transition is between states with non-bonding character, here between carbon and nitrogen non-bonding orbitals (for the radical-carbazole systems). The second is where there is electron-hole separation to vibrationally-decoupled molecular fragments. We consider that this second mechanism is operative symmetry-broken triplet and singlet excited states shown in Figure 1(a).

Since the ground state exchange energy is low (~1.2K), we expect a 3:1 ratio of triplet:singlet ground states down to low temperatures. We observe however that the ratio of triplet to singlet PL is closer to 1:1 (noting that the overall PLQE is > 90%), and consider this must arise through significant ISC from triplet to singlet. The TA evolution shown in Figure 2(c) shows this happens beyond 100 ps. We assign the thermally activated delayed emission at 640 nm to rISC, from the $S_1$ to the $T_1$ state. These observations of competing singlet and triplet excitation and de-excitation pathways, and evidence for intersystem crossing, ISC and reverse ISC, rISC sets up scope for photoexcited spin polarisation in the ground state.

We carried out ground and excited state electron spin resonance (ESR) measurements, on 0.1wt% doped polystyrene thin-films containing $M_2TTM$-3Flr-$M_2TTM$ (for 50µM toluene solutions, see SI). In the dark, continuous wave (CW) ESR measurements at 298 K, Figure 4(a), shows clear evidence for triplets with full field $|\Delta m_s|$ = 1 transitions which could be simulated with D = 30.3 MHz and E = 1.6 MHz. Further confirmation of a ground state triplet state is given by the observation of a $|\Delta m_s|$ = 2 transition at half field.

Ground state pulsed ESR was performed, and the resulting echo-detected field sweep shows two peaks, due to dipolar coupling in the triplet with additional structure from g anisotropy, see figure 4(b). Echo decay measurements at the two peak positions behave similarly; a bi-exponential fit to peak 2 gives $T_m$ of 280 ns and 830 ns. The longer $T_m$ of 830 ns is comparable to that of a perchlorotriphenylmethyl (PTM) monoradical(*34*). Inversion recovery experiments show that the $T_1$ (spin-lattice) relaxation time also has a fast and slow component of 2.5 and 8.0 μs. As $T_m$ is clearly not limited by $T_1$, creating a nuclear free environment could further increase the $T_m$.

One critical component for molecular systems to be utilised in quantum information applications is the ability coherently manipulate the spin state(*35*). Rabi nutation experiments shown in Figure 4(c), show the ability of this system to be coherently driven between two states with multiple oscillations seen at room temperature.

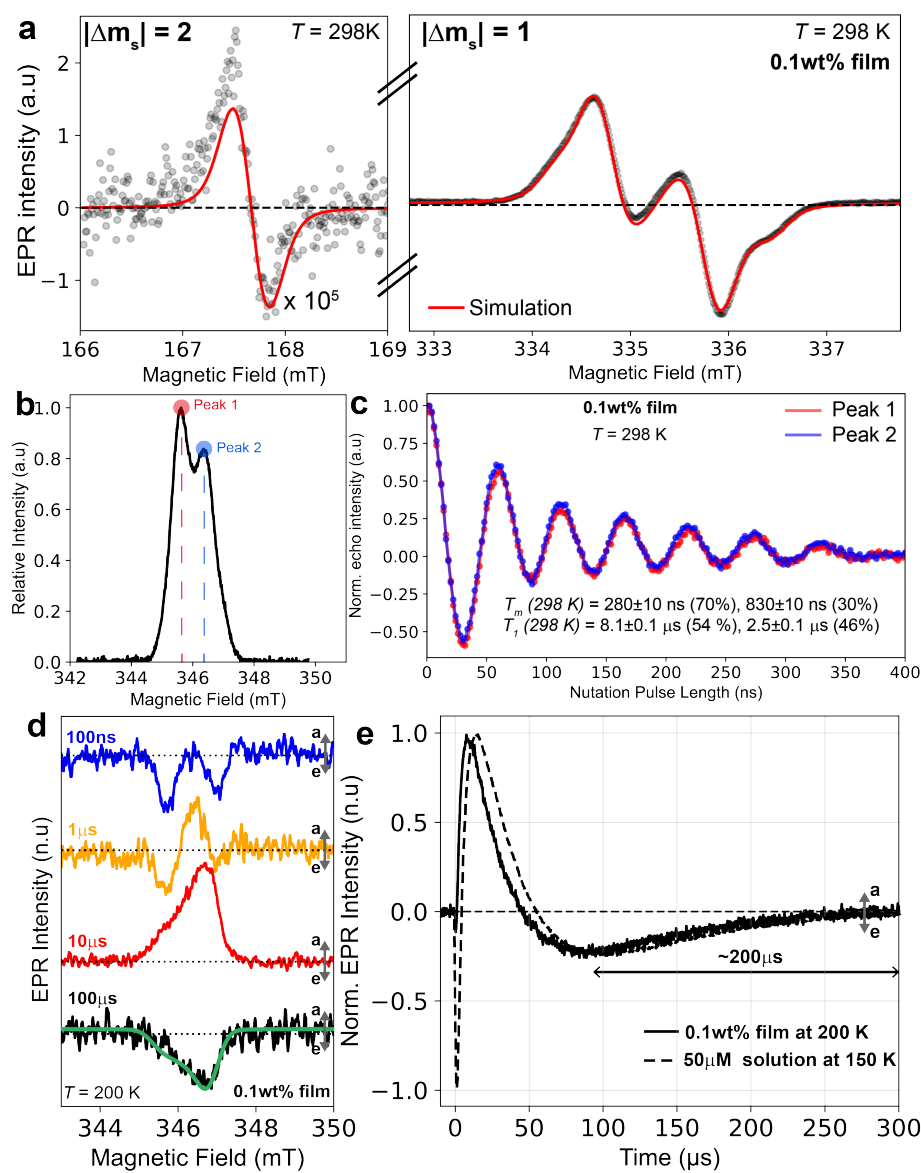

*Figure 4: Photoinduced long-lived ground-state spin polarisation. **(a)** Half-field(left) and Full-field(right), dark, continuous wave X-Band ESR spectra at 298K which show the |Δm$_s$|= 2 transition in a 0.1wt% doped polystyrene film (black dots) and the simulation (red line) of all features as a triplet species. **(b)** the EDFS spectrum of the diradical. **(c)** Rabi oscillations on the main pulsed ESR transitions, labelled as Peak-1 and Peak-2 in (b), of the diradical at 345.62 mT and 346.67*

*mT as a function of nutation pulse length. Further measurements of $T_1$ and $T_m$ on the films at room temperatures are used to derive the quoted values in the figure, the inversion-recovery and phase-memory measurements are shown in the SI.* **(d)** *Transient ESR spectral slices derived at the beginning of each temporal decade ($10^2$ ns – $10^5$ ns) of a 0.1wt% doped polystyrene film of $M_2TTM$-3Flr-$M_2TTM$ diradical at 200K. The spectra are obtained at the quoted time point after a 532nm laser excitation lasting for 5 ns which repeats at a frequency of 100 Hz. The data highlights long-lived photogenerated spin-polarized ESR signals persisting beyond 200μs. The simulation of the 100μs slice (green solid line) confirms polarisation in $T_+$.* **(e)** *Kinetic traces of the TrESR signal in the 345 mT – 348 mT region for the 0.1wt% doped polystyrene film at 200 K(black-solid lines) and the 50 μM frozen toluene solution at 150 K(black-dashed lines) of the $M_2TTM$-3Flr-$M_2TTM$ diradical.*

Transient ESR (TrESR) was measured with 532 nm 5 ns pulsed photoexcitation. Figure 4(d) shows spectra measured at different times after excitation. At the timescales shown here, we consider the dominant response to arise from non-thermal spin sublevel populations in the electronic ground state, noting that the triplet luminescence response has an initial decay time of 10 ns. At 100 ns and 1 μs two emissive features are observed, which we consider arise from sublevel transitions in the triplet state, with preferential population in the $T_0$ and $T_+$ sublevels. This indicates either preferential filling of the $T_0/T_+$ or depopulation of the $T_-$ spin-levels, the latter via ISC in the excited state to the singlet with spin-sublevel selective transition probabilities. At 10μs, the spectra show a broad absorptive feature and beyond 50 μs this ESR feature reverses sign to become emissive, and is measurable beyond 200 μs, see Figure 4(d). A similar trEPR response is found in frozen solution spectra at 150 K, with the only difference being the initial emissive feature to much more pronounced (See Figure SI 32-33 for 2D maps). Note that the 5 ns pulse excitation (100 Hz repetition rate) is shorter than the PL lifetime and precludes spin pumping via re-excitation. A possible mechanism for the sign reversal (Figure 4(c)) is spin sub-level selective excited state depopulation(*3, 36*). The later time sign reversal shown in Figure 4(c,d) is very unusual, and we consider requires spin repopulation from an excited state reservoir(*37*). A possible mechanism to provide this re-population channel is by rISC from the reservoir of long-lived excited state singlets which persists to these timescales at 200 K (Figure 2(c) and S11). We confirm the long-lived photo-generated spin polarisation beyond 50 μs to be due to overpopulation of the ground state $T_+$ sublevel by simulation of the late-time trESR spectra where we require a ground state triplet population distribution of $p(T_+) > p(T_0) = p(T_{-1})$ but with the same D and E parameters as for the cwEPR dark spectrum.

Ground state spin polarisation mediated by a spin-optical interface can be revealed in photoluminescence-detected magnetic resonance, PLDMR, measurements. We explore a range of PLDMR conditions in Figure 5. The Zeeman energy exceeds D above 2.8 mT, so at *X*-band and *D*-band, the triplet levels are set by the applied $B_{\|z}$ fields, setting up ISC and rISC to underpopulate principally the $T_-$ at the longer times responsible for the PLDMR (see Figures 4(c,d)). At 100 K, and X-band ESR conditions, shown in Figure 5(a), we see that the PLDMR recorded over the full PL spectrum shows a positive contrast $>10^{-3}$ %. We report low temperature PLDMR under *D*-band conditions in Figures 5(b-c). The PLDMR transition is large with contrast ~4% and is rather narrow with a FWHM of 2 mT, reminiscent of the dark-cwESR spectrum of the diradical. The size of the PLDMR response is relatively agnostic to microwave power but scales with sample magnetisation, as seen in Figure 5(c), tracking the PL response in Figure 3(b). This demonstrates that the response scales with the population of ground state triplets. We resolve PLDMR throughout the full PL spectrum, shown in Figure 5(d), at the resonance point. We find that applying the microwaves bleaches triplet emission

(near 640 nm) and enhances singlet emission (near 700 nm) reaching values of *ΔPL/PL* up to 8% at 300 mK. This behaviour arises due to preferential ISC from the triplet T*$_0$ excited state to S$_1$, state as shown in Figure 5(e) which is induced due to the ground state T$_{-1}$ →T$_0$ transition induced by the microwave (See SI section V for more details). The mechanistic insights for the quantum mechanical origins of PLDMR responses from alternant diradicals has also been recently and independently explored by Poh *et al*(*17*).

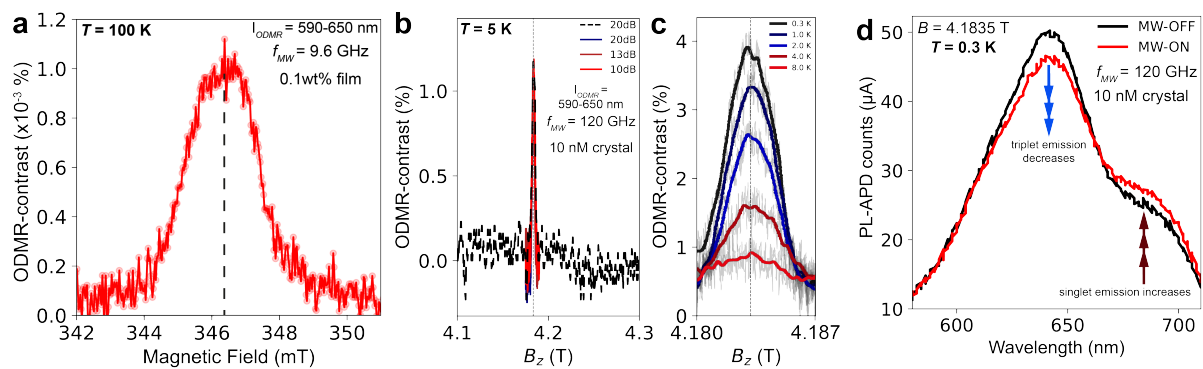

*Figure 5: **Spin-Optical interface in diradicals.** We probe the spin-optical interface using probed using Photoluminescence detected magnetic resonance spectra.  **(a)** at X-band ESR conditions (9.7 GHz, 0.346 T) ΔPL/PL >10$^{-3}$ % at 100 K, and **(b)** at D band (120 GHz, 4.1T) with different microwave powers. In both cases a large magnitude sharp transition is observed optically. (c) Temperature dependence of the D-band PLDMR shows a ΔPL/PL > 7% at temperatures between 300 mK and 8 K. **(d)** Shows spectrally resolved PLDMR under conditions at the PLDMR resonance point at 4.1835 T.  All PL detected in (a)-(c) and utilised a long pass filter (>420 nm) to cut-off the laser line. Samples for (a) were thin-films made with 0.1wt% M$_2$TTM-3Flr-M$_2$TTM doped into Polystyrene, in (b) and (c) we doped 10nM of M$_2$TTM-3Flr-M$_2$TTM into PhCl$_3$ crystal, the crystals prepared using slow evaporation of a solution which were subsequently washed and polished are described in SI.section XIII.*

In summary, we have reported open shell organic diradical semiconductors with close to unity PL yields, from both triplet and singlet excitons.  This molecule can function as a spin-optical interface with the capability to provide access for ground state manipulation and exhibits long-lived spin polarisation at relatively high temperatures.  We have shown that this meets the criteria for a viable spin-optical interface(*35*): *(i)* A spin-sublevel selective excited state process can be used to prepare the ground state spin population after photo-excitation preferably with long spin-coherence times of at least 10*x* the optical lifetimes(*38*), *(ii)* The ground spin-state can be manipulated using magnetic fields or coherently using microwaves,  and *(iii)* Luminescence can be obtained from the electronic transitions within the different ground spin-states so that we can optically read-out the prepared spin-state. We demonstrate agreement with all the criteria by showing that: *(i)* We can optically prepare and coherently control a spin polarised ground state which persists beyond 200 μs at 200 K; *(ii)* We show the existence of both magneto-optical and spin-optical interfaces by demonstrating control over the ground state population in the spin-singlet and spin-triplet states by applying either a magnetic field or by coherently driving the ground-state using microwaves; *(iii)* We are able to read out the finally prepared state using PL spectroscopy. For magnetic field control, established using magneto-photoluminescence, MPL, we obtain contrasts of up to 50%. In the case of microwave control, established via photoluminescence detected magnetic resonance, PLDMR, we observe very large optical contrasts of ~8% in both singlet and triplet emission.

This approach to the magnetic control over the ground state spin and photogeneration, as well as optical readout of spin-polarisation in the ground state, is relevant to the operational principles of optical qubits such as single photon emitting high spin defects in *c*-Si, diamond or *h*-BN. In contrast to such defects, where placement and thus large scale entanglement is difficult(*39*) because there is no direct control of defect placement, there is real scope for designed open shell, highly luminescent, carbon-based molecular spin-optical systems, as exemplified here for the $M_2$TTM-3Flr-$M_2$TTM diradical. We consider the advances reported here provide a strong and unexpected basis for further developments for optically-controlled qubit platforms.

## Acknowledgements

We acknowledge Sebastian Gorgon, Lucy Walker, Jeannine Grüne, Biwen Li, Lujo Matasovic for helpful discussions. We thank Timothy J. H. Hele for discussions on the quantum chemical modelling. We thank Sam Bayliss and Hannah Stern for discussions on optically detected magnetic resonance (PLDMR) experiments. **R.H.F**, **P.M.** and **R.C.** received funding from the European Research Council under the European Union's Horizon 2020 research and innovation programme (Grant Agreement No. SCORS – 101020167). **R.C** also was supported by the European Union's Horizon 2020 project for funding under its research and innovation programme through Marie Skłodowska-Curie Actions (Grant Agreement No. 859752, HEL4CHIROLED). **P.M** also acknowledges support from the European Union's Horizon 2020 research and innovation programme (Grant Agreement No. 891167, PROLED). **H.A.B** acknowledges support from the Engineering Physical Sciences Research Council (EPSRC, grant number EP/S003126/1). **A.C, Y.S** and **M.M** acknowledges support from ANR-20-CE92-0041 (MARS), IDF-DIM SIRTEQ, and the European Research Council (ERC) under the European Union's Horizon 2020 research and innovation programme (grant Ballistop agreement no. 833350). **J.B and N.A.P** acknowledge support from the German Research Foundation (DFG grant number BE 5126/6-1). **P.G** thanks the Cambridge Trust and the George and Lilian Schiff Foundation for a PhD scholarship and St John's College, Cambridge, for additional support. **Y.B** thanks the Winton Programme for Physics of Sustainability for funding. **Y.F** thanks financial support by the Engineering Physical Sciences Research Council (EPSRC, grant number EP/W017091/1) Programme Grant. **T.B.E.M** acknowledges financial support from Cambridge NanoDTC and the Winton Programme for Physics of Sustainability for funding. **D.C** acknowledges the Herchel Smith fund for an early career fellowship. **A.R** received funding from the European Research Council under the European Union's Horizon 2020 research and innovation programme (Grant Agreement No. 758826). Research in Mons is supported by the Belgian National Fund for Scientific Research (FRS-FNRS) within the Consortium des Équipements de Calcul Intensif – CÉCI (grant number U.G.018.18), and by the Walloon Region (LUCIA Tier-1 supercomputer; grant number 1910247). **D.B.** is a FNRS research director. **J.C** acknowledges to European Union for its Marie Curie Individual fellowship (HORIZON-MSCA-2022-PF-01-01, project nº 101106941).

## Author Contributions

**R.H.F** and **H.A.B** conceived the project. **P.M** designed, synthesized the molecules, characterised the synthesized molecules, assessed their purity and carried out the cyclic voltammetry studies. **R.C** developed the project, planned and conducted the steady-state spectroscopy, transient spectroscopy and cryo-magneto-optical experiments. **O.M** and **D.C** synthesized the 3,6-dibromo-9,9-dioctyl-9*H*-fluorene monomer. **N.A.P.** performed the CW, Pulsed, Transient ESR and PLDMR experiments, analysed the results and discussed results with **J.B**. **Y.S**, **V.D**, and **M.M** helped to perform and interpret the cryo-magneto-optical experiments under the supervision of **A.C**. **P.G** performed ultrafast fs-transient absorption measurements. **Y.B** and **S.J.W** performed Density Functional Theory calculations. **Y.F** and **T.M** performed additional cw ESR experiments. **R.H.F, R.C, H.A.B** and **P.M.** wrote the manuscript with contributions from all other authors.

## Corresponding Authors:

**Richard H. Friend** (rhf10@cam.ac.uk), **Hugo A. Bronstein** (hab60@cam.ac.uk), **Alexei Chepelianskii** (alexei-chepelianskii@universite-paris-saclay.fr)